\documentclass{ws-procs9x6}

\begin{document}

\title{Dark Energy and Non--Linear Perturbations}

\author{C. van de Bruck}

\address{Department of Applied Mathematics\\
The University of Sheffield \\
Hounsfield Road \\
Sheffield S3 2RH\\
United Kingdom}

\author{D.~F. Mota}

\address{Astrophysics Department, University of Oxford\\
Keble Road, OX1 3RH, United Kingdom \\
and \\
Institute of Theoretical Astrophysics, University of Oslo \\
N-0315 Oslo, Norway}

\maketitle
\abstracts{
Dark energy might have an influence on the formation of 
non--linear structures during the cosmic history. For example, 
in models in which dark energy couples to dark matter, it  
will be non--homogeneous and will influence
on the collapse of a dark matter overdensity. We use the 
spherical collapse model to estimate how much influence dark 
energy might have.}

\section{Introduction}
One of the most important goals of contemporary cosmology is to unreveal 
the properties of dark energy. This energy form is thought to be 
responsible for the observed accelerating expansion of the present day 
universe. There are several methods used to study the properties of 
dark energy. The important ones make use of the anisotropies in the 
Cosmic Microwave Background Radiation (CMB), the evolution of 
large scale structure formation (LSS) and/or the distances of 
high redshift supernovae (see e.g. the overview by Peebles and Ratra\cite{overview}). 

In this contribution we address the question whether dark energy 
can have some impact on the formation of non--linear structures 
in the universe, such as clusters of galaxies or galaxies itself. 
In doing so, we assume that dark energy is a scalar field, pervading 
the universe. It will obviously depend on the properties
of dark energy if this scalar field has any influence on non--linear 
structure formation. For example, if dark energy couples to dark matter, 
there is an extra force between the dark matter particles, mediated 
by dark energy. This possibility was first discussed in detail by Wetterich\cite{Wetterich} and 
afterwards in particular by Amendola\cite{Amendola}.
However, even if dark energy does not couple to 
dark matter, backreaction effects of the gravitational field might 
influence the evolution of dark energy inside a non--linear overdensity. 

Here, we use the spherical collapse model to study the influence of dark 
energy on non--linear structure formation.

\begin{figure}[ht!] 
\centerline{\epsfxsize=4.1in\epsfbox{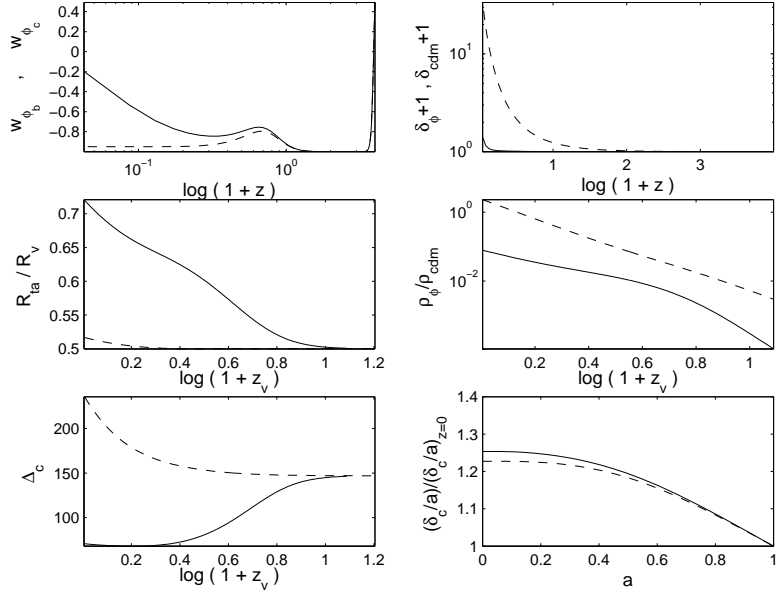}}   
\caption{Double exponential potential (see text). Top left panel: Full collapse of dark energy: 
evolution of $w_\phi$ in the background (dashed line) and 
inside the overdensity (solid line) as a function of redshift.
Top right panel: evolution of $\rho_{\phi,{\rm overdensity}}/\rho_{\phi,{\rm outside}}$ 
(solid line) and $\rho_{\rm cdm, overdensity}/\rho_{\rm cdm, outside}$
as a function of redshift in the case of clustering of dark energy. Middle left panel: 
$R_v/R_t$ as a function of virialisation redshift in the case of homogeneous dark energy 
(dashed line) and collapsing dark energy (solid line). Middle right panel: The ratio 
$\rho_\phi/\rho_{dark matter}$ inside the overdensity as a function of virialisation 
redshift. Bottom left: $\Delta_c$ as a function of virialisation redshift in the case
of homogeneous (dashed line) and inhomogeneous (solid line) dark energy. Bottom right:
the linear density contrast as a function of the scale factor $a$. 
\label{copeland}}
\end{figure}

\section{The spherical collapse model} 
The spherical collapse model is based on the assumption that an 
overdensity can be treated as a homogeneous and isotropic, closed 
``sub-universe'', embedded in our universe. For a cold dark matter
universe, this assumption is justified by Birkhoff's theorem. However, 
as soon there is a second fluid, such as radiation, dark energy, etc., 
this (idealised) over-density can, and will, exchange energy with its surroundings. 
In the case of dark energy studied here, however, the exchange will only 
be a small fraction of the total energy, since dark energy is subdominant 
for most parts of the cosmic history. Therefore, it should be a not 
too bad description once the energy out- or inflow is specified. However, 
the spherical collapse model does not specify the energy outflow (which we denote
by $\Gamma$) of dark energy into the surroundings of the overdensity. Therefore, 
we have to make assumptions about $\Gamma$. In our work\cite{us} we considered two extreme cases. 
In the first case, we assumed that dark energy does not cluster at all but is 
homogeneous throughout space. In the second case we assumed that it fully collapses
along with dark matter. Clearly, the reality might be somewhere between these two 
possibilities. However, we will get an idea about the difference to be expected.  

The typical time-evolution of a spherical overdensity is as follows:
Initially, the overdensity expands with almost the same rate as the 
universe. However, since the density is higher, the expansion of the 
overdensity will eventually slow down until it starts to contract. 
The point at which the expansion turns into contraction is called 
{\it turnaround}. Without dissipation, the overdensity would collapse
to a singularity. However, in reality, energies inside the overdensities
viralise and the sphere ends up at some final radius (virialisation). 
In our calculations, we follow the evolution of the sphere until 
virialisation is reached. We have studied several different 
potentials and refer to our paper\cite{us} for the details of the calculation. 
Here, we consider only three different potentials as models for dark energy. 
The first one is a double exponential potential\cite{copeland}. The form is 
$V(\phi) = M(\exp(\alpha\phi) + \exp(\beta\phi))$. The second is the well known
supergravity potential\cite{brax}, which is $V(\phi) = M\exp(\phi^2)/\phi^\gamma$.
The third is a exponential potential with power--law modifications\cite{kostas}, i.e.
$V(\phi)= M(A+(\phi-B)^2)\exp(-\gamma\phi)$. 

\begin{figure}[ht!] 
\centerline{\epsfxsize=4.1in\epsfbox{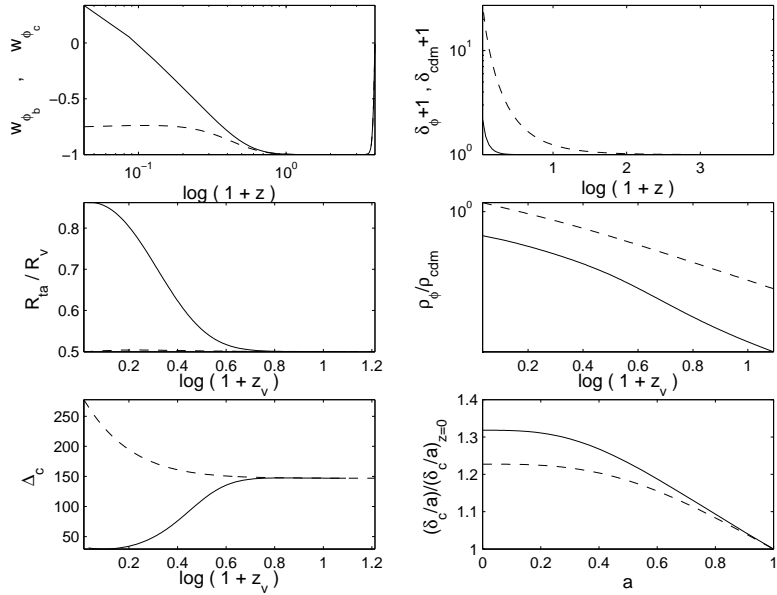}}   
\caption{The same as Figure 1, but for the supergravity potential (see text). \label{brax}}
\end{figure}

The results of our calculations can be found in Figs.1 -- 3. Apart 
from the equation of state of dark energy inside and outside the 
overdensity and the density contrast of dark energy and dark matter, 
we calculate the ratio of the radius at turnaround $R_t$ to the radius at 
virialisation $R_v$ as a function of redshift at which the overdensity 
virialises ($z_v$). The latter is equal to 0.5 in the case of the 
standard cold dark matter model and has a slight dependence on the 
cosmological constant in a $\Lambda$CDM model \cite{rees}. It can be seen from 
Figs. 1, 2 and 3 that in the case of a homogeneous dark energy component 
the ratio $R_t/R_v$ depends on the model of dark energy, but is still of 
order 0.5. If dark energy collapses together with dark matter, this 
quantity depends strongly on the virialisation redshift. As a result, 
the density contrast at the time of virialisation 
$\Delta_c=\rho_{\rm cdm, inside}(z_v)/\rho_{\rm cdm, outside}(z_v)$ becomes
strongly dependent on $z_v$ as well. 

\begin{figure}[ht!] 
\centerline{\epsfxsize=4.1in\epsfbox{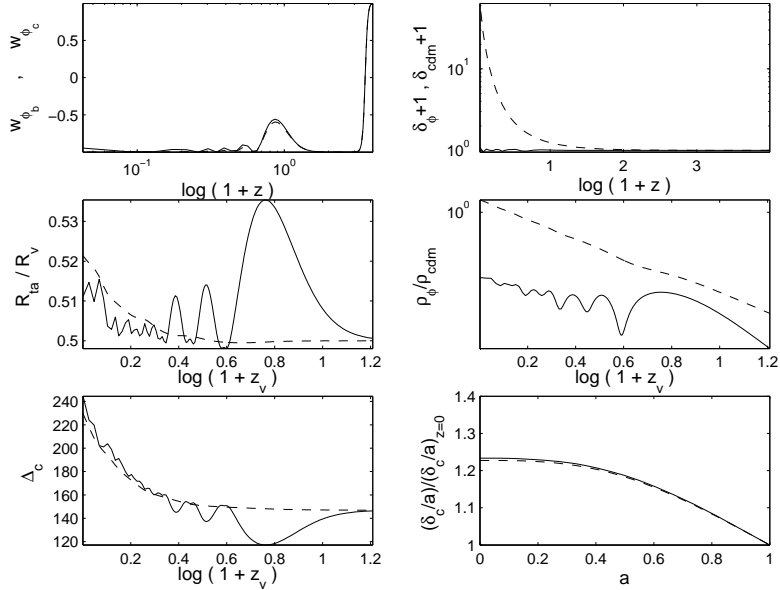}}   
\caption{The same as Figure 1 but for the modified exponential potential (see text). \label{scordis}}
\end{figure}

On the other hand, the ratio of the energy densities of dark matter and dark 
energy depends on the clustering properties of dark energy, but is small at 
the typical redshift of cluster formation, even if dark energy clusters. 

We have also checked the dependence on our assumptions. For example, a more conservative 
assumption would be that dark energy only clusters {\it after} turnaround, i.e. once the 
overdensity is decoupled from the rest of the universe. Before that, the field is homogeneous. 
In this case, there is still a big difference from the homogeneous case (see our work\cite{us} 
for more details).

\section{Outlook}
The spherical collapse model susggests, that dark energy can have 
an important impact on non--linear structure formation, even if it 
is dynamically unimportant for most of the time during the cosmic
history. We have considered two extreme cases, namely that 
dark energy either fully collapses together with dark matter or that
dark energy is homogeneous throughout space. Its interesting to note
that even {\it if} dark energy collapses with dark matter, it will 
stay in the linear regime (or sometimes in the quasi non-linear regime, 
in which the density contrast is of order 1), whereas dark matter
is in the highly non--linear regime. In fact, although dark matter enters
a highly non--linear regime, the dark energy density contrast deviates 
from unity only slightly.  

The aim of our approach was not to make predictions for structure 
formation, but rather to investigate if dark energy can at all 
have a significant impact on the details of structure formation. 
Clearly, our work shows that it can have, but the answer depends strongly 
on the details of the theory. Our work is rather limited, since the 
spherical collapse model can not predict how much dark energy will 
flow out of the overdensity. For this, a fully relativistic approach 
has to be taken, in order to calculate the amount exactly. However, our 
work clearly indicates that the spherical collapse model has to be used
with care, when comparing models of dark energy with data. 

On the other hand, our work raises some questions for future work: even if 
dark energy clusters strongly, the differences to a theory where it does not
cluster should not be too large, since at the redshift of structure formation 
(i.e. $z \geq 3$) the differences between the theories are small. How will we 
be able to differentiate between the theories using only structure formation data? 
Note that the case of a dark energy which collapses with dark matter is rather extreme, 
so its likely that any signal of dark energy clustering will be even smaller, once the exact 
value of $\Gamma$ is known. 

\section*{Acknowledgements}
C.v.d.B. was supported by PPARC. D.F.M. is supported by Funda cao Ciencia e a Tecnologia.

\end{document}